\documentclass[useAMS,usenatbib]{mn2e}
\usepackage{times}
\usepackage{epsfig}
\usepackage{amssymb}
\usepackage{amsmath} 
\usepackage{aas_macros}
\title[Automated line fitting algorithm]
{ALFA: an automated line fitting algorithm}
\author[R. Wesson]
{R. Wesson$^1$ \\
$^1$European Southern Observatory, Alonso de C\'ordova 3107, Casilla 19001, Santiago, Chile\\
}
\date{Received:}

\begin{document}
\maketitle

\begin{abstract}

I present the Automated Line Fitting Algorithm, {\sc alfa}, a new code which can fit emission line spectra of arbitrary wavelength coverage and resolution, fully automatically.  In contrast to traditional emission line fitting methods which require the identification of spectral features suspected to be emission lines, {\sc alfa} instead uses a list of lines which are expected to be present to construct a synthetic spectrum.  The parameters used to construct the synthetic spectrum are optimised by means of a genetic algorithm.  Uncertainties are estimated using the noise structure of the residuals.

An emission line spectrum containing several hundred lines can be fitted in a few seconds using a single processor of a typical contemporary desktop or laptop PC.  I show that the results are in excellent agreement with those measured manually for a number of spectra.  Where discrepancies exist, the manually measured fluxes are found to be less accurate than those returned by {\sc alfa}.

Together with the code {\sc neat} (Wesson et al. 2012), {\sc alfa} provides a powerful way to rapidly extract physical information from observations, an increasingly vital function in the era of highly multiplexed spectroscopy.  The two codes can deliver a reliable and comprehensive analysis of very large datasets in a few hours with little or no user interaction.

\end{abstract}
 

\section{Introduction}

Knowledge of the abundances in ionised gases in the universe is of crucial importance in a variety of astrophysical contexts.  The abundances measured in planetary nebulae (PNe) or Wolf-Rayet (WR) ejecta nebulae provide constraints on theories of stellar nucleosynthesis and evolution (e.g. \citealt{2009ApJ...690.1130K};  \citealt{2011arXiv1110.1186M}; \citealt{1992A&A...264..105M}; \citealt{2011MNRAS.418.2532S}), while the abundances in Galactic and extragalactic H~{\sc ii} regions provide insights into the current composition of the interstellar medium (ISM) and therefore are vital constraints for the output of galactic chemical evolution models (e.g. \citealt{1997nceg.book.....P}; \citealt{2003ceg..book.....M}).

Obtaining estimates of chemical abundances from observations of an ionised nebula requires the measurement of the fluxes of emission lines, with various ratios subsequently used to infer the physical conditions and abundances of elements in the gas.  The more observations one has, the more important it is to reliably automate the measurement of line fluxes, and cutting edge astronomical instrumentation can now deliver data in extremely large quantities; for example, MUSE on the Very Large Telescope at Paranal in Chile delivers 90,000 spectra simultaneously for each observation, covering wavelengths from 4800 to 9300\,{\AA} at a resolution of $\lambda/\Delta\lambda$=3000.  A night of MUSE observations can easily yield a million spectra (\citealt{2014Msngr.157...13B}).  In contrast, a night observing with a non-multiplexed spectrograph such as UVES on the VLT might only yield 10--20 spectra.

Highly multiplexed spectroscopy allows the investigation of astronomical objects in unprecedented and extraordinary detail.  However, to exploit the data fully, the extraction of information from it needs to be made extremely efficient.  If each spectrum obtained in a MUSE observation were to contain only 20 lines, and if the measurement of the flux of a single line, including all preparation and user input, were to take one second, then the line measurement process for a single observation would take just under three weeks.  In practice, the number of lines detected could easily be a factor of 10 higher.

While computer codes to optimise the parameters of spectral fits have been available for a long time, the selection of initial fit parameters, the number of functions to fit to a spectrum, and the continuum to subtract have generally needed to be selected by the user.  The process can be extremely tedious and time consuming.  The line fluxes presented in Appendix 1 of \citet{2005MNRAS.362..424W} took me approximately five working days to measure, and the experience of measuring and identifying lines at a rate of only 90 per hour for a week contributed to me leaving astronomy and pursuing an alternative career for some two and a half years following the submission of the paper.  The original idea for a tool like {\sc alfa} was conceived around this time.

A highly automated means of measuring emission line fluxes is thus desirable.  Such a means could involve searching a spectrum for features which appear to be emission lines and fitting Gaussian functions to them.  However, a number of difficulties arise -- in low or intermediate resolution spectra, many lines may be blended, and deriving meaningful information from an unresolved blend requires knowledge of which lines should be present.  Noise may be confused with spectral features, such that either valuable processing time is wasted fitting a Gaussian function to a spurious feature, or that a real feature is wrongly flagged as noise and ignored.  The algorithm for discriminating between noise and real spectral features is very likely to introduce significant uncertainties into the measurements at low signal to noise (\citealt{1994A&A...287..676R}; Wesson et al. 2016).  And even assuming that all the lines present in the spectrum can be reliably measured, they must still then be identified.  Some semi-automated means of doing this have been developed, such as \textit{EMILI} (\citealt{2003ApJS..149..157S}), but their use is still impractical when considering the workload involved in analysing tens of thousands of spectra in a single observation.

{\sc alfa} takes a different and, as far as I know, hitherto untried approach.  Rather than aiming to detect emission line features, measure them and then see which known spectral line each feature corresponds to, the code starts by assuming the emission lines which will be present in the spectrum being analysed.  It then assumes that the lines will have a Gaussian profile, the width of which is determined by the spectrograph resolution, and optimises the parameters for all lines in the spectrum by means of a genetic algorithm.  The code is designed to be fast, and the chosen approach has a number of advantages over alternative methods.

\section{Algorithm}

\subsection{Input file formats}

{\sc alfa} is written in Fortran 95.  It reads files in either plain text or FITS format.  A plain text file is assumed to contain a list of wavelengths and fluxes.  When FITS files are read in, the behaviour of the code depends on the dimensionality; a 1D FITS file is treated in the same way as a plain text file.  For 3D files, the code assumes and x and y are spatial dimensions with the z-axis being wavelength.  It extracts and measures the spectrum contained in each individual pixel.  Thus, for example, a reduced {\sc muse} data cube can be given directly to the code, and it will return the line flux measurements for every pixel in the data cube.

\subsection{Continuum subtraction}

Before emission line fluxes can be measured, the continuum must be subtracted.  Manual analyses typically fit a continuum to manageable sections of the spectrum, either by eye or by fitting a smooth function to regions of the spectrum identified as being pure continuum.  A disadvantage of fitting in sections is that normally no constraint is applied to ensure that the sections join up.  Unphysical jumps in the estimated continuum flux are thus likely to exist, introducing a probably small but none the less unquantified systematic uncertainty into the measured fluxes.

{\sc alfa}'s continuum fitting algorithm operates globally to avoid this potential pitfall.  The code analyses the input spectrum and calculates the 25th percentile of the flux values in a moving window 100 data points wide.  This approach leaves 50 points at the beginning and end of the spectrum with no estimated continuum flux; the code fills in these values using the 26th and (n-26)th points, for which there is an estimated flux.

\begin{figure*}
\includegraphics[width=0.47\textwidth]{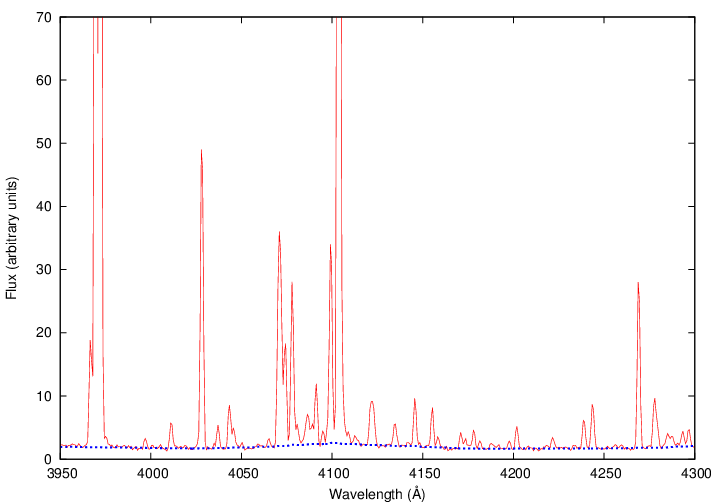}
\includegraphics[width=0.47\textwidth]{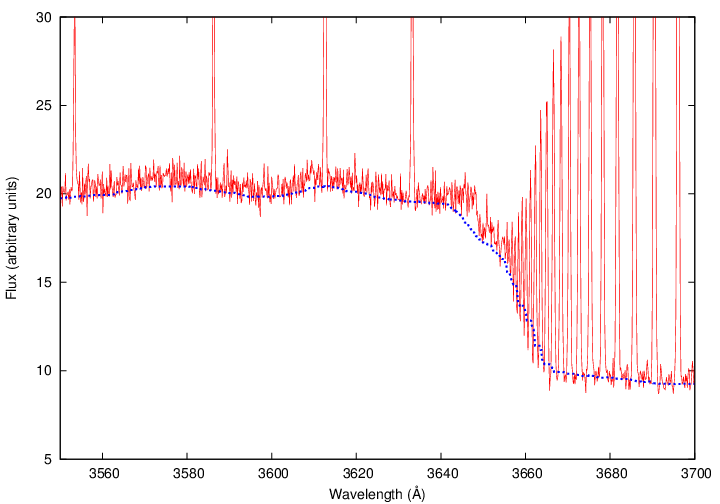}
\caption{Illustration of {\sc alfa}'s continuum fitting for FORS2 observations of NGC 6778 presented in \citet{Jones21012016} (l) and high resolution X-SHOOTER observations of Tc-1 (Aleman et al., in prep.) in the region of the Balmer jump (r).  The observed spectrum is in red and the estimated continuum is in blue.}
\label{continuumfits}
\end{figure*}

Figure~\ref{continuumfits} shows some examples of continua estimated using this method.  In all the spectra I have fitted with {\sc alfa} so far, visual inspection suggests that this approach has fitted a reasonable continuum.  In particular, it gives a reasonable fit to the continuum in the region of jumps, such as the hydrogen Balmer and Paschen jumps.  The code reports the magnitude of these jumps for use in later calculations.  In regions with few or no emission lines, this method will tend to underestimate the actual continuum flux, which would be better approximated by the median of the flux values in the case of a pure linear continuum plus Gaussian noise.  Generally the effect on line flux measurements will be very small, but in case some alteration of the window size and percentile is necessary to improve the continuum fit, the default values can be overridden by the user.

\subsection{Emission line fitting: genetic algorithm}

Traditional manual emission line measurement relies on the selection by eye of features which appear to be emission lines, aided in the case of blends by prior information about which lines may be present in the blend.  Once features are measured, line identification is carried out as a subsequent step.  The process can be extremely time consuming, especially for deep spectra.  Anecdotal evidence suggests that for spectra containing several hundred lines, many months may be required to carry out a careful and complete measurement and identification.  The process can be extremely tedious, and may thus be prone to errors.  In addition, routines which fit Gaussian functions to emission line profiles generally struggle to fit more than about 20 simultaneously.

{\sc alfa} takes an alternative approach.  In the case of emission line spectra, it is relatively easy to compile a list of emission lines that might be present.  {\sc alfa} takes such a list as input and then constructs a synthetic spectrum consisting of Gaussian profiles for every line in the list.  The parameters of the Gaussian functions are then optimised using a genetic algorithm, which I describe below.

\subsubsection{Genetic algorithms}

Genetic algorithms are a class of optimisation algorithm in which candidate solutions undergo mutation and selection to evolve towards better solutions.  They were first suggested as early as 1950 by Alan Turing (\citealt{TURING01101950}), and use the principles of evolutionary biology to find solutions to complex problems.  In particular, mutation and natural selection are used to efficiently explore very large parameter spaces.  The process by which a solution is reached can be divided into several stages.

\subsubsection{Creation of initial population}

The first stage of a genetic algorithm is to create the initial ``population".  For {\sc alfa}, each member of the population is a synthetic spectrum, which is represented as having a single spectral resolution, a single redshift, and for each line in the spectrum, a wavelength and a peak flux.  30 such synthetic spectra are created, with arbitrary peak fluxes at wavelengths taken from the reference line list.  The initial guess for the velocity is zero, and that for the resolution is determined from the wavelength sampling assuming Nyquist sampling, but these values can also be specified by the user.  To reduce the number of calculations required by the code, the flux of each synthetic line is calculated only within 5$\sigma$ of the line centre.  The error introduced by this approximation is negligible, with the fraction of the flux lost for each line being 5.7$\times$10$^{-7}$.

\subsubsection{Goodness of fit}

For every member of the population, the goodness of fit is calculated.  {\sc alfa} does this by calculating a synthetic spectrum from all of the individual line parameters.  The sum of the squares of the differences between the synthetic spectrum and the observed spectrum is then calculated, and used as a measurement of the goodness of fit.

\subsubsection{Breeding}

{\sc alfa} ranks the population according to the goodness of fit, and if the code has reached the final generation, then the best fitting member of the population is returned as the result of the fit.  Otherwise, the code discards all but the best performing 30\% of the population.  The best fitting member of the population is retained unaltered in the following generation, while to generate the rest of the population, random pairs of the survivors are chosen, and their line parameters averaged to create members of the new generation.  The breeding proceeds until the population reaches 30 members once again.

\subsubsection{Mutation}

Once 29 offspring have been generated, {\sc alfa} introduces mutations into their genetic code (the best fitting member of the previous generation is excluded from the mutation process).  For the resolution, and for every line peak, a random number $r$ between 0 and 1 is chosen, and the parameter multiplied by a function m(r) defined as follows:

\begin{equation}
m(r) = 
\begin{cases}
\frac{r}{0.05} & (r<0.05) \\
1.0 & (0.05<r<0.95) \\
2+\frac{r-1}{0.05} & (r>0.95) \\
\end{cases}
\end{equation}

The application of the mutation function thus leaves 90\% of parameters unchanged, with the remaining ten per cent multipled by a number between zero and two.  The redshift is multiplied by (999+m(r))/1000, such that the mutation function is between 0.999 and 1.001, equivalent to a variation of $\pm$ 300\,km\,s$^{-1}$.  This restriction ensures a more efficient exploration of the redshift parameter space than would result from the use of the unrestricted mutation function.

\subsubsection{Evolution}

The entire process is then repeated with the new generation.  {\sc alfa} evolves the population through 500 generations, a value chosen conservatively and which ensures that a good fit is achieved.  200 generations is found generally to work sufficiently well for most purposes, and thus the code can be modified to run a factor of almost 2.5 times quicker if required.

\subsubsection{Choice of algorithm parameters}

The number of generations, the population size, the mutation function and the fraction of the population selected for breeding in each generation are each somewhat arbitratrarily chosen; studies have not found any generally applicable rules for the choice of these parameters (for example, \citealt{Ochoa00optimalmutation}, \citealt{neumuller2012}), and I instead experimented with each parameter to try to optimise the speed of the algorithm.  The most time consuming computation is the generation of the synthetic spectrum, and so the lower the product of population size and number of generations, the quicker the code runs.  After initially selecting the best performing half of each generation for breeding, I found that reducing this fraction gave a more rapid convergence.  The mutation function was chosen to be computationally very simple, and to allow mutations to occur over a continuous range from very small to large.  The mutation rate is constant over all generations.

\subsubsection{{\sc alfa}'s full approach}

After initially coding {\sc alfa} to fit all lines in the spectrum simultaneously, I subsequently found that it was far quicker to fit the spectrum in smaller sections, and the performance of the algorithm was also better.  When fitting the entire spectrum at once, weak lines were sometimes poorly fitted as the criterion for selection is dominated by the uncertainties associated with stronger lines, such that a bad fit to a weak line has only a small penalty associated with it.  In addition, minor uncertainties in wavelength calibrations can lead to significant systematic shifts of line centres over large wavelength ranges.  Spectrograph resolutions may also not be constant with wavelength.

{\sc alfa} therefore uses a two pass approach to fitting entire spectra.  First of all, using a small subset of the brightest lines, it fits the whole spectrum to determine the approximate velocity and resolution .  Then, in sections each of 400 data points, it uses those initial guesses to fit all the lines in the section.  In the first pass, the velocity is allowed to vary by $\pm$900\,km\,s$^{-1}$ and the resolution by a factor of two from its input value; in the second pass, they are allowed to vary by $\pm$60\,km\,s$^{-1}$ and $\pm$500 respectively.  The optimal values obtained in each section are passed to the next section as initial guesses.

This two pass approach is fast and robust, and allows for variations in the spectral resolution and the quality of the wavelength calibration across the full spectral range.  Each section overlaps with the preceding and succeeding sections, but only lines whose peaks lie within the unique part of the section are fitted.  Thus, for lines lying at the edge of a section, their complete profile is used to optimise the Gaussian fit, but no line is fitted more than once.

\begin{figure}
\includegraphics[width=0.47\textwidth]{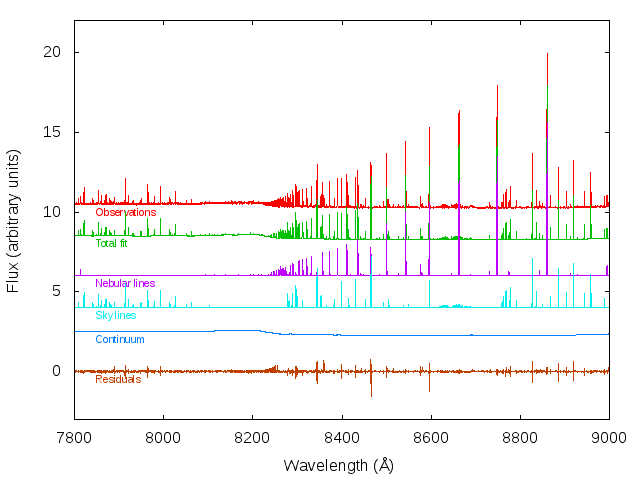}
\caption{Example of a fit to a region of a spectrum containing a continuum jump as well as numerous sky and nebular emission lines.  The components of the fit are offset for clarity.}
\label{components}
\end{figure}

Before fitting nebular emission lines, {\sc alfa} can also remove telluric emission lines, for cases where sky emission was not measured separately.  It does this in exactly the same way as the nebular fitting, but with the line of sight velocity only allowed to vary between -0.5 and +0.5\,km\,s$^{-1}$.  The list of sky lines included with {\sc alfa} is a subset of 438 of the lines listed in the catalogue of \citet{2003A&A...407.1157H}.  An example of an {\sc alfa} fit to a spectrum in a region containing a continuum jump as well as numerous nebular and sky emission lines is shown in Figure~\ref{components}.

\subsection{Estimation of uncertainty}

Once the line parameter fitting has converged, {\sc alfa} estimates the uncertainty on the measured line fluxes using the residuals obtained by subtracting the best fitting solution from the continuum-subtracted input spectrum.  In a moving window 20 units wide, the two highest residuals are discarded, and the root mean square of the remaining points is calculated.  The highest residuals are discarded because they generally do not reflect the actual noise in the spectrum but can instead be large residuals from strong lines, which would otherwise cause a spuriously high uncertainty to be estimated.  The ratio of the peak flux of each line to the RMS of the residuals at the wavelength of the line is then taken to be the signal to noise ratio at the centre of the line.

The code then applies the empirical relation described by \citet{1992PASP..104.1104L} for estimating the uncertainty on a line fit based on the signal to noise ratio at the centre of a line.  The relation is

\begin{equation}
  \frac{F}{\sigma_F} = 0.67\left(\frac{FWHM}{\Delta\lambda}\right)^{1/2}\left(\frac{f_o}{\sigma_f}\right)
\end{equation}

where F is the integrated line flux, $\Delta\lambda$ is the wavelength sampling interval, and f$_0$ is the peak flux.  \citet{1992PASP..104.1104L} found this relation to be applicable independent of the noise model assumed, whether pure poissonian or with such deviations from poissonian as are found in typical astronomical instruments.

Once the uncertainties are calculated, the significance of the fitted lines can be determined.  {\sc alfa} then removes from the output all lines which for which the signal to noise ratio of the flux measurement is less than 3.0.  For the remaining lines, their identifications and fluxes are written to two files, one a \LaTeX-formatted table for easy inclusion in publications, and the second a plain text file suitable for direct input into the Nebular Empirical Analysis Tool (\citealt{2012MNRAS.422.3516W}).  A third file containing the original spectrum, the fitted spectrum, the estimated continuum, the continuum-subtracted original spectrum, the sky spectrum if calculated, and the residuals of the fit is also created.

\subsection{Line blends}

In deep intermediate resolution spectra, many lines can be blended.  Common instances of relevance to problems in nebular astrophysics include the blends of O~{\sc ii} and [S~{\sc ii}] lines at 4068--4075{\AA}, and of O~{\sc ii} and C~{\sc iii} lines at $\sim$4650{\AA}.  {\sc alfa} ignores blends until after the fitting has been completed.  At this point, it examines the line list for any lines separated by less than the half-width at half-maximum of the lines at the calculated resolution.  These lines are flagged as blended, and all flux is then attributed to one member of the blend.  Blend flagging takes place before uncertainty estimation.  The calculation of uncertainty is based on the RMS in the residuals in a moving window centred on each line, containing 20 data points.  In a Nyquist-sampled spectrum, blends would be separated by less than 2 data points, so that the uncertainty calculation does not significantly differ over the wavelength range of the blend.  An uncertainty based on the total flux in the blend and the RMS in a 20-unit window containing the blend is thus appropriate.

\subsection{Unfitted lines}

Lines which are not in {\sc alfa}'s line catalogue will not be fitted.  The catalogue I have compiled for use with {\sc alfa} is intended to be thorough and deep but it cannot be truly comprehensive, and a limited selection of lines is necessary to avoid over-fitting the data.  It can therefore happen that {\sc alfa} fails to fit lines which are present in spectra being analysed but absent from its line catalogue.  To aid in the identification of such lines, the code calculates the average difference between each data point in the spectrum and its two neighbours, and where the wavelength lies more than 5$\sigma$ away from any line already fitted, the highest values of the average difference are reported so that the user can inspect the spectrum to see if any unfitted lines are indeed present.

\section{Performance}

I assessed the performance of the code in several ways, to check various aspects of its performance.  These aspects were the speed of the code and the factors affecting it; how well the code reproduced previously published results; how reliably the code fitted the same spectrum when run many times; and how the flux measurements made by {\sc alfa} compare to those made for the same spectrum by a number of other line fitting routines.

\subsection{Speed}

Keeping the run time as short as possible was one of the fundamental aims in creating {\sc alfa}, and so every effort has been made to optimise the {\sc fortran} code for speed.  The run time depends on several factors, some intrinsic to the code and others intrinsic to the spectrum being analysed.  Within the code, the number of generations and the size of the population are the primary determinants of how fast it runs, and I selected values of these which minimised the time to achieve reliable fits.  The number of data points in the spectrum being analysed, and the number of lines which {\sc alfa} attempts to fit, both affect the time taken to run the code, which scales approximately linearly with both factors.  The default line catalogue contains 686 emission lines between the atmospheric cut-off and the near-infrared, and in testing on a number of spectra actually containing 100-200 emission lines, fits take 5-10 seconds on a single processor of a four year old 32 bit desktop computer.  {\sc alfa} is parallelised using OpenMP (\citealt{dagum1998openmp}), so that when dealing with data cubes, as many spectra can be fitted simultaneously as there are processors available.  The time taken to read even very large FITS files into memory is small - a few seconds for a MUSE data cube containing 90,000 spectra with $\sim$3500 data points each, for examples.  I recently analysed a 5.1 Gb data cube obtained in the science verification run for MUSE, using 2 processors of a 32 bit desktop computer; {\sc alfa} took about 20 hours to analyse the cube, during which time 41,022 pixels were analysed and just over 2 million emission lines were fitted.

\subsection{Internal consistency - 1000 fits to the same spectrum with {\sc alfa}}

\begin{figure*}
\includegraphics[width=0.97\textwidth]{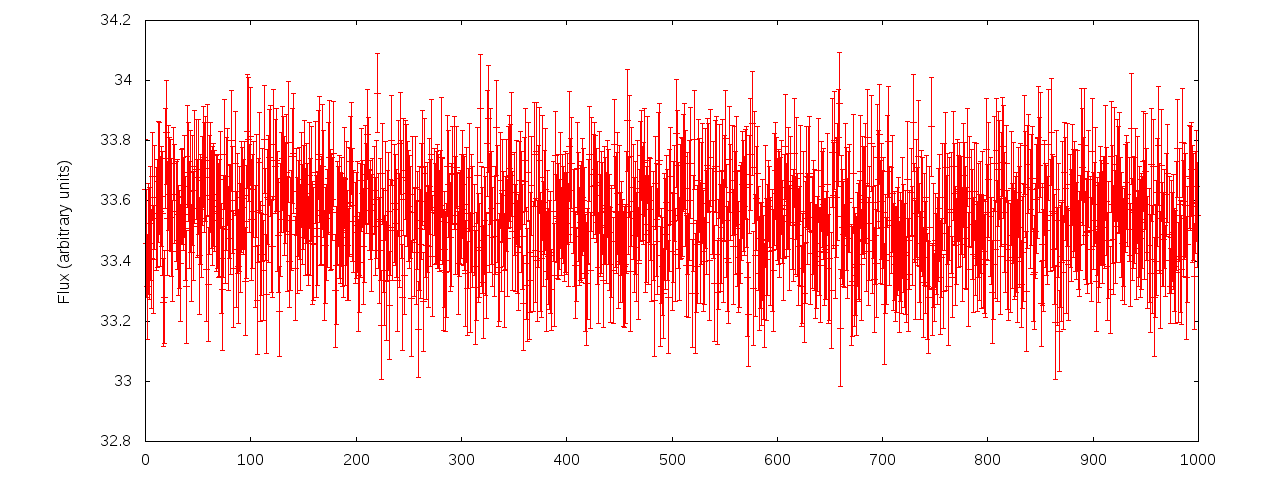}
\includegraphics[width=0.97\textwidth]{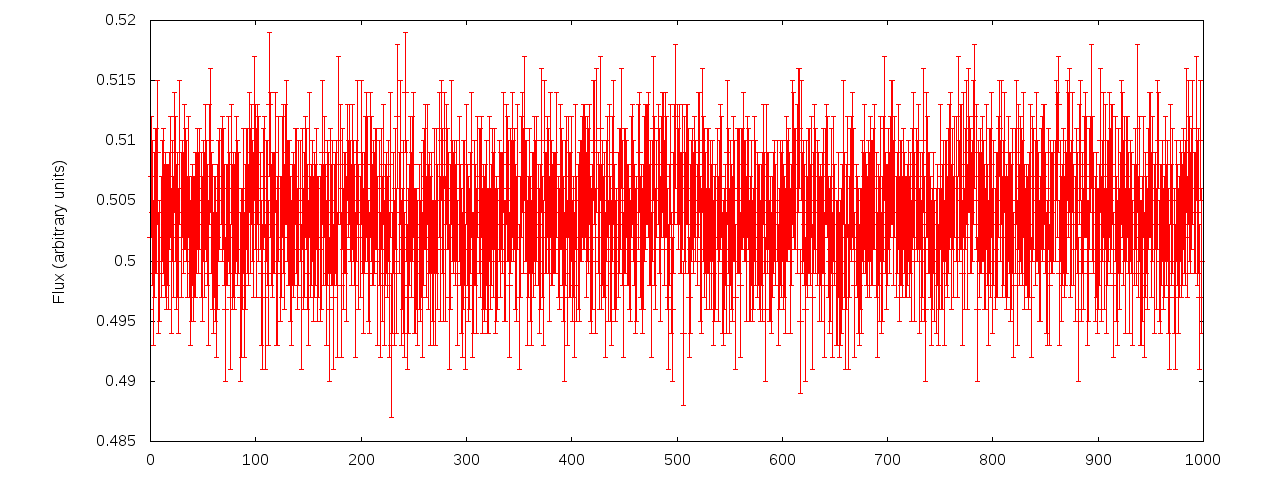}
\includegraphics[width=0.97\textwidth]{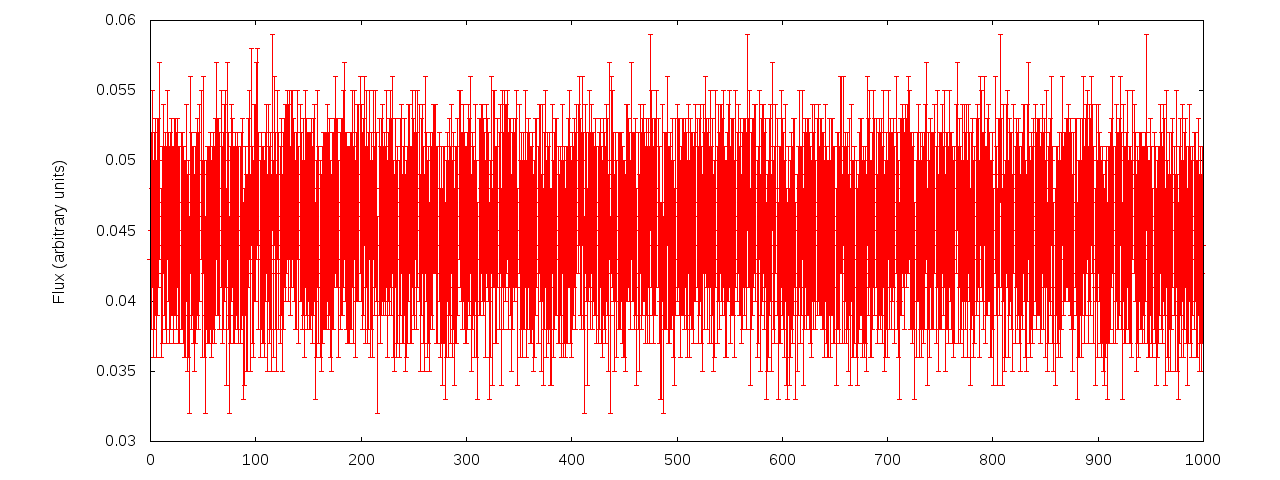}
\caption{1000 flux measurements of lines of varying strengths: [O~{\sc iii}] 4959{\AA} (top), He~{\sc i} 4713{\AA} (middle), N~{\sc iii} 4634{\AA} (bottom)}
\label{internalconsistency}
\end{figure*}

A genetic algorithm is inherently probabilistic, and so {\sc alfa} does not report exactly the same line fluxes each time it is run on the same spectrum.  For the code to be useful, the line fluxes reported should be the same to within their estimated uncertainties, and so I ran {\sc alfa} 1000 times on the same input spectrum to check that this was the case.  Figure~\ref{internalconsistency} shows the ranked sets of reported line fluxes for a selection of lines.  The distribution of reported line fluxes within each set of 1000 should correspond to the uncertainty estimated from the RMS in the residuals, and Figure~\ref{internalconsistency} shows this to be the case.

From the 1000 runs, I also checked whether line flux pairs were correlated with each other for marginally resolved lines.  {\sc alfa}'s criterion for considering lines resolved is that the separation of their line centres should exceed the locally calculated half width at half maximum.  If the line fluxes of marginally resolved lines are indepently measured then they should be uncorrelated; an anticorrelation of fluxes would be observed if the line fluxes cannot be calculated independently.  In the spectrum I used for the test, there were two O~{\sc ii} lines separated by 0.74{\AA}, where the FWHM of the lines was 0.71{\AA}; the fluxes reported by {\sc alfa} showed no correlation.

\subsection{External consistency - reproducing previous results}

\begin{figure*}
\includegraphics[width=0.47\textwidth]{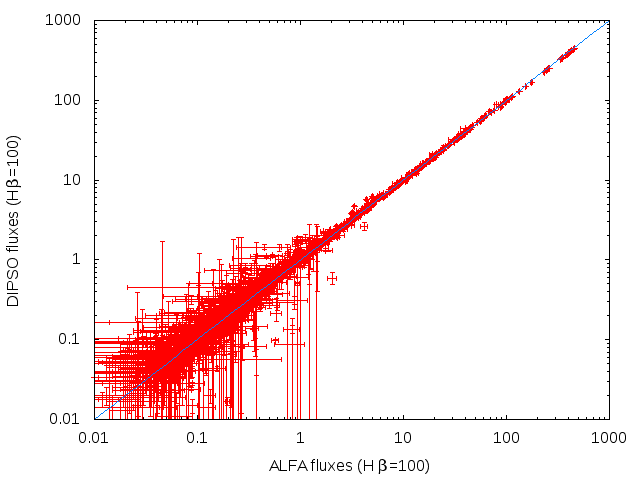}
\includegraphics[width=0.47\textwidth]{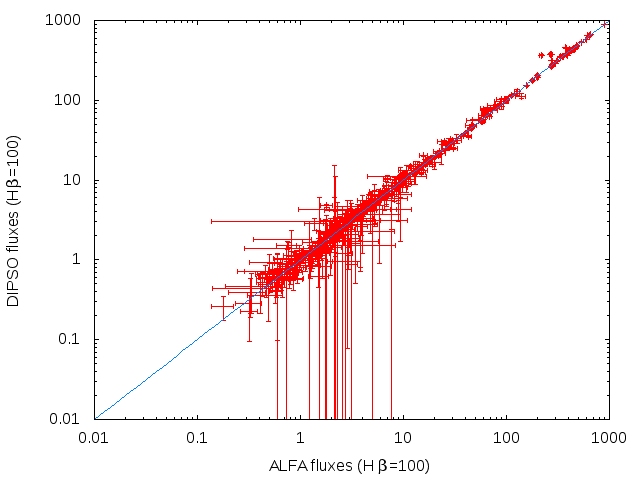}
\caption{Comparison between line fluxes measured using {\sc alfa}, and those previously measured using {\sc dipso} and published in \citet{2005MNRAS.362..424W}, for high resolution blue spectra (l) and low resolution red spectra (r)}
\label{wessoncomparison}
\end{figure*}

I used {\sc alfa} to measure line fluxes in the spectra originally presented in \citet{2005MNRAS.362..424W}.  As mentioned earlier, these were measured over the course of approximately a week, with line identification then being carried out in a semi-automated way by reference to line lists in earlier papers presenting deep spectroscopy of planetary nebulae.  {\sc alfa} fitted the 46 spectra of 23 objects in around 7 minutes on a four year old 32-bit desktop computer, and about three minutes on a two year old 64 bit laptop, both running the Linux Mint operating system, and with the code compiled using gfortran.  From the line lists produced by {\sc alfa} and those presented in \citet{2005MNRAS.362..424W}, I extracted the common lines measured, and compared their reported fluxes.  \citet{2005MNRAS.362..424W} observed their target nebulae in two instrument configurations - a high resolution ``blue" spectrum covering 3500--5000 {\AA} and a low resolution ``red" spectrum covering 3800--8000{\AA}.  The comparison between {\sc alfa} and manual fluxes for both sets of spectra is shown in Figure~\ref{wessoncomparison}, and shows excellent agreement between the pairs of fluxes, which are almost all the same to within their reported uncertainties.  Discrepancies between the fluxes could mean that {\sc alfa} is wrong, or that the earlier manual fluxes were wrong.  A few outlying points are visible, and in every case that I investigated, these are blended lines, which were not reported as such in \citet{2005MNRAS.362..424W}.

\subsection{External consistency - comparison to other line fitting algorithms}

In this section I compare the performance of {\sc alfa} to other widely used line fitting algorithms.  On 7 May 2015, I used the ``Astronomers" group on facebook to ask a large sample of professional astronomers what tools they use to measure emission line fluxes.  239 responses were received, including some people who reported using several codes.  86 people reporting using codes of their own creation, while 77 reported using NOAO's {\sc iraf} suite.  The other programs people reported using were {\sc class}, {\sc dipso}, XSpec, {\sc casa}, {\sc gandalf}, Splat, Sherpa, {\sc midas} and {\sc isis}, with between 2 and 21 users each.  

To assess how well {\sc alfa} performed relative to the other software packages available for the purpose of emission line fitting, I later posted another request to the same group, in which I provided a sample of a spectrum of a planetary nebula, containing approximately 30 emission lines, and asked people willing to join the experiment to measure whatever lines they considered to be present and report their positions and fluxes.  The aim of the experiment was twofold; for strong unblended lines, the results would show how well each of the codes performed, with the expectation being that for such lines, the reported fluxes and uncertainties would be extremely similar.  For fainter and blended lines, the results would also reveal the influence of subjectivity on emission line fitting.  Astronomers not familiar with the particular wavelength range or particular type of spectrum being analysed might neglect faint lines on the margins of detectability which other astronomers would consider worth fitting.

There were 8 responses to this survey, which was sufficient for comparisons to be made between the results.  The codes used were {\sc splat} (2 users), {\sc iraf} (2 users), Filili (one user), {\sc dipso} (one user), {\sc isis} (one user) and {\sc alfa} (one user).  The number of lines fitted ranged from 7 to 45, with a median of 18.  The reported time taken to measure the lines was between 10 seconds and 4 hours, with a median of 12 minutes.  The number of lines identified seems to be somewhat dependent on the code being used: the two {\sc iraf} users reported the fewest lines, the two {\sc splat} users the third and fourth most lines, and the other three codes the most lines.  The users of {\sc alfa}, {\sc dipso}, {\sc isis} and {\sc splat} reported measurement uncertainties, while the other three did not.

\begin{figure*}
\includegraphics[width=0.97\textwidth]{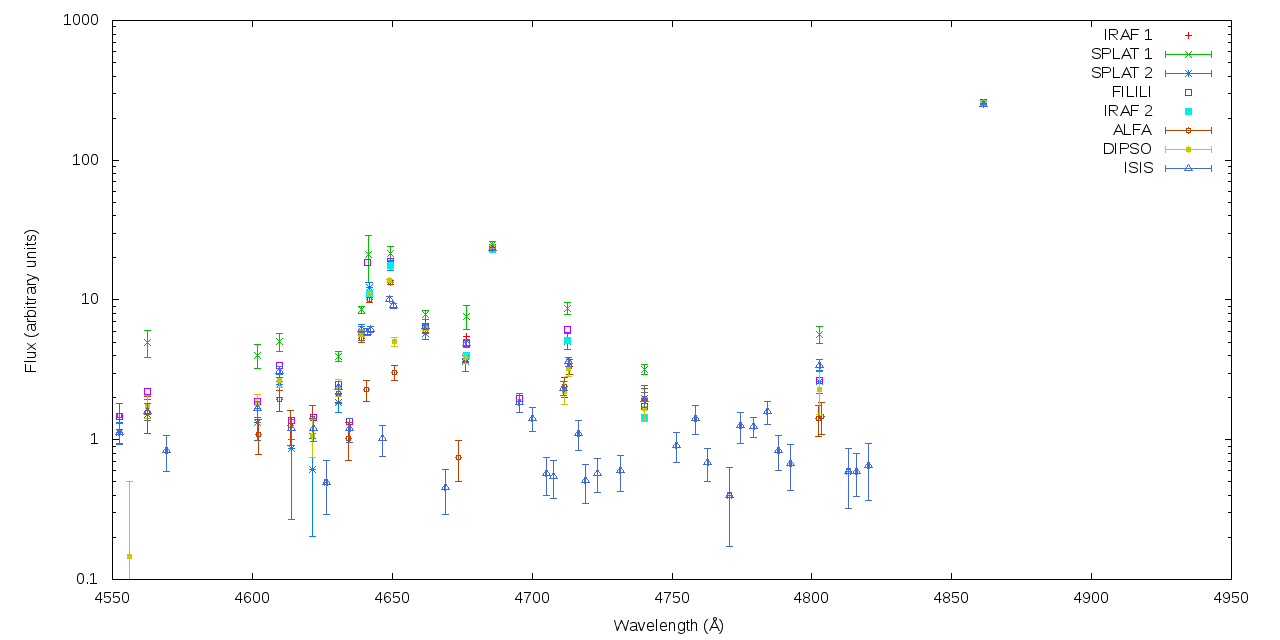}
\caption{Flux measurements from seven different users analysing the same spectrum.}
\label{fittingtest}
\end{figure*}

Remarkably, there were no lines for which all measurements were completely consistent -- all estimates lying within the uncertainties of all the other measurements.  In all cases, at least one flux measurement lay outside the reported uncertainties of at least one other flux measurement.  All eight sets of measurements are plotted in Figure~\ref{fittingtest}.  While there was at least broad agreement in the fluxes measured for obvious and unblended lines, the results diverged significantly for blended and weak lines.  For example, the feature in the spectrum at 4712{\AA} is a blend of an [Ar~{\sc iv}] line and an He~{\sc i} line; {\sc alfa} considers the lines marginally resolved and reports fluxes for both, but most participants in the survey reported only one flux for the feature.  With the complex blend of recombination lines at 4650{\AA}, most participants did not attempt to identify the components of the blend; obviously in a small and informal test such as this, it would have been unreasonable to expect them to spend much time doing so, and so they reported only the integrated flux of each feature.

A further illustration of the subjectivity of line flux measurements comes from the two pairs of users using the same code.  While the two {\sc iraf} users agreed very well except for one line where the fluxes differed by 30\%, the two {\sc splat} users reported fluxes which systematically differed from each other, by 2-3 units.  This corresponds approximately to the continuum flux integrated over a typical line width, and so I suspect that one user did not subtract the continuum.

Five users reported uncertainties on their flux measurements; these uncertainties showed poor agreement, with large differences in the estimates in most cases.

The magnitude of the differences between the 8 sets of measurements was surprisingly large, and highlights the need for objective and reproducible means of measuring line fluxes.  {\sc alfa} is intended to be exactly that, and its performance in comparison to the other 5 codes was good.  It identified the most lines, and a careful inspection of the input spectrum suggests that there were neither false positives nor false negatives.  Most of the other codes had false negatives, and {\sc isis} reported some false positives.  {\sc alfa} also took the least time, requiring no action other than typing a single command in a terminal window.  On the small amount of data used in the test, it took less than a second to fit all the lines it detected.  Finally, it resolved several blends reported as single features by other users, and correctly flagged other features as blends in its output line list.

\section{Discussion}

I have described a new code, {\sc alfa}, which employs a novel methodology to rapidly fit emission line spectra with minimal user interaction.  The high degree of automation facilitates the extraction of information from extremely large data sets.  {\sc alfa} can measure all the emission lines in deep spectra in a few seconds, and is parallelised so that when analysing data cubes, multiple pixels can be fitted simultaneously.  The fitting is shown to be very reliable, with an excellent match to line fluxes measured manually.  Though the code is inherently probabilistic and will never return exactly the same flux measurements twice, 1000 fits to the same spectra show that the reproducibility of the fits is excellent.  The objectivity and speed of the fitting is highly desirable in the era of highly multiplexed spectroscopy and the generation of very large numbers of spectra.

I briefly discuss here some possible applications of {\sc alfa} beyond those I initially designed it for.  {\sc alfa} was designed in the first place to measure emission lines with Gaussian profiles in optical spectra.  However, the methodology is highly generalisable and applicable to many other problems of extracting information from spectra.  In the first place, other wavelength ranges are trivially fittable - the only requirement is the preparation of a catalogue of lines suitable for the wavelength range.  Secondly, the instrumental line function is easily changed to other forms.  One example of a more complex form is that of Fourier transform spectrographs, such as SPIRE on the Herschel Space Observatory, for which the instrumental line profile is a sinc function.  The secondary lobes of the sinc function can make it difficult to see by eye which lines are present.  One technique commonly used is to apply an apodising function to the data before the fourier transform.  This results in spectra with approximately Gaussian line profiles, at the expense of resolution.  Widely used apodising functions have a roughly linear relation between the normalised FWHM of the final line profile, and the logarithmic reduction of intensity of the largest secondary lobe: a 90\% reduction results in a roughly 30\% increase in the FWHM, while a 99\% reduction results in a 60\% increase in FWHM (\citealt{Naylor:07}).  Thus, if apodisation can be avoided, more information is retained.  The genetic approach of {\sc alfa} trivialises the localisation and measurement of emission lines in unapodised spectra. \citet{2010A&A...518L.144W} presented fluxes of emission lines measured from apodized Herschel-SPIRE observations; I tested a version of {\sc alfa} modified to fit sinc profiles on the unapodized spectra, and measured the same fluxes to within the reported uncertainties.

One disadvantage of fitting sinc profiles is that, while a Gaussian profile can be truncated at $\pm$5$\sigma$ without introducing any significant uncertainty, a sinc profile extends essentially across the whole spectrum.  {\sc alfa} thus runs considerably slower when fitting sinc profiles compared to Gaussian profiles, and the fitting took many minutes compared to a few seconds to fit large numbers of Gaussian profiles. 

With some further modification, {\sc alfa} could also be used to fit multiple velocity components, which may be useful for H~{\sc ii} regions, and planetary nebulae for which the expansion velocity is larger than the instrumental resolution.  The genetic approach would be able to optimise the number of components without user input.

Further applications are no doubt possible.  The code is freely available to facilitate its easy adaptation by anyone who may be interested in applying it to different problems.  It is licensed under the GNU General Public License, and can be obtained from http://www.github.com/rwesson/alfa.

\section{Acknowledgments}

I thank C. Sabiu for interesting discussions which inspired me to consider the application of genetic algorithms to emission line spectroscopy; D. Stock and P. Scicluna for sharing a great deal of {\sc fortran} knowledge; M. Leal Ferreira, I. Aleman, J. Cami and D. Stock for testing and critiquing the code and giving development ideas; and D. Jones, M. Nowak, H. Daniels, I. Aleman, H. Gunther and M. Crnogorcevic for carrying out fits for the purposes of comparing codes.  Finally I am grateful to the anonymous referee for their helpful review which improved both the code and this paper.  This work was co-funded under the Marie Curie Actions of the European Commission (FP7-COFUND).

\bibliographystyle{mn2e}
\bibliography{alfa}

\end{document}